\documentstyle[12pt]{article}
\begin{document}
\thispagestyle{empty}
\begin{raggedleft}
UR-1401\\
ER-40425-261\\
hep-th/9502042\\
Dec.\ 1994\\
\end{raggedleft}
$\phantom{x}$\vskip 0.618cm\par
{\huge \begin{center}
FADDEEV-JACKIW APPROACH TO HIDDEN SYMMETRIES\footnote{This work is
supported by CNPq, Bras\'\i lia, Brasil}
\end{center}}\par
\vfill
\begin{center}
$\phantom{X}$\\
{\Large Clovis Wotzasek\footnote{Permanent address:
Instituto de F\'\i sica,
Universidade Federal do Rio de Janeiro,
Brasil}\footnote{E-mail:clovis@urhep.pas.rochester.edu} }\\[3ex] {\em
Department of Physics and Astronomy\\ University of Rochester\\ Rochester,
NY 14627 \\ USA\\ email: clovis@urhep.pas.rochester.edu }
\end{center}\par
\vfill
\begin{abstract}

\noindent The study of hidden symmetries within Dirac's formalism does not
possess a systematic procedure due to the lack of first-class constraints
to act as symmetry generators.  On the other hand, in the Faddeev-Jackiw
approach, gauge and reparametrization symmetries are generated by the null
eigenvectors of the sympletic matrix and not by constraints, suggesting
the possibility of dealing systematically with hidden symmetries throughout
this formalism.  It is shown in this paper that indeed hidden symmetries
of noninvariant or gauge fixed systems are equally well described by null
eigenvectors of the sympletic matrix, just as the explicit invariances.
The Faddeev-Jackiw approach therefore provide a systematic algorithm for
treating all sorts of symmetries in an unified way.  This technique is
illustrated here by the SL(2,R) affine Lie algebra of the 2-D
induced gravity proposed by Polyakov, which is a hidden symmetry in the
canonical approach of constrained systems via Dirac's method, after
conformal and reparametrization invariances have been fixed.

\end{abstract}
\vfill
\newpage

\section{Introduction}

Invariant theories, both of gauge and reparametrization type, have been
dealt with in the last 50 years by Dirac's Hamiltonian
formalism\cite{Dirac}, with its categorization of constraints as first and
second-class and the separation of equalities as weak and strong.  Within
this context the symmetries of a given theory are generated by a complete
set of first-class constraints obtained by a well defined algorithm, and
these constraints are used to reduce the original Hilbert space into the
subspace of physical solutions.  Nevertheless, noninvariant theories and
gauge-fixed invariant models presenting only second-class constraints may
still possess a set of (hidden) symmetries that cannot be unveiled by
Dirac's methodology.  As we know, second-class constraints are used to
alter the original (Poisson) bracket operation so that they can be imposed
as operators identities, but cannot be used as symmetry generators.
Within Dirac's formalism, the disclosure of those symmetries that may be
(possibly) hidden in a model will depend then crucially on our ability and
experience on the subject, due to the lack of a systematic methodology.

Recently, Faddeev and Jackiw\cite{fadjack} proposed an alternative approach
to constrained systems that avoids the separation of
constraints into first and second-class and the use of weak and strong
equalities.  This new method of analysis has been successfully used by many
authors\cite{noga,ovid,jun,edgardo,mkk,marcelo}, and is by now a standard
technique to deal with constrained systems. In a series of papers
\cite{bw1,bw2,mw,montani}, this author and
collaborators have proposed an algorithm, based on the Faddeev-Jackiw
technique, to treat systems with constraints, both invariant and
noninvariant, on the same foot: for noninvariant systems the sympletic
matrix defining the geometrical structure of the theory will always end up
being invertible even if it was not to begin with.  This matrix inverse
will then provide the Dirac brackets used in the quantization process(up to
ordering ambiguities).  For
gauge theories, the sympletic matrix is never invertible, but it has been
shown \cite{mw,montani} that the null eigenvectors of this matrix are the
generators of the intrinsic gauge symmetry of the model.

It is our intention in the present work to show that those symmetries that
persist after gauge fixing and are called hidden in the literature due to
the lack of a set of first-class constraints to generate them, need no
special treatment in the Faddeev-Jackiw context, being described by a set
of null eigenvectors of the sympletic matrix exactly as for the explicit
symmetries.  While for those systems where Dirac's formalism is free of
pathologies, the advantage of Faddeev-Jackiw's method over Dirac's seems
to be only of practical nature (it is simpler and quicker to use), the
case of hidden symmetries shows a definite edge in favor of the
Faddeev-Jackiw geometric approach.

To illustrate this systematic process of finding the hidden symmetries of
an arbitrary action, we study the SL(2,R) symmetry of Polyakov's 2D
induced gravity.  The present status of this theory is as follows.  The
SL(2,R) current algebra symmetry of the induced gravity was discovered by
Polyakov \cite{polyakov} a few years ago.  Later on this symmetry has led
to an exact solution of the 2-D quantum theory of gravity interacting with
conformal matter \cite{kpz}.  It allows us to write down recursion
relations for the (Euclidean) Green functions and to determine the
renormalization of the parameters of the theory.  This symmetry, which is
hidden under the more conventional approach to current algebra, was
discovered by Polyakov by analyzing the anomaly equation.  In spite of its
importance for the development of this very active field, the physical
origin of this symmetry remained obscure\footnote{See note added at the
end of the paper.} (for a nice review on the development of this theory
and a complete list of references see Ref.\cite{rothe}).

The study of this symmetry under the canonical point of view has been
pursued by many authors \cite{abdalla,em,jose,barcelos}, mostly based
on Dirac's theory of constrained systems\cite{Dirac}.  The puzzling point, as
indicated by Egorian and Manvelian\cite{em} and by Barcelos-Neto
\cite{barcelos}, is that the local
form of Polyakov's model in the light cone gauge has only second-class
constraints when written in Dirac's light-front coordinates, and no
constraints whatsoever if written in instant-front variables\cite{dirac2}.
Dirac's
method for dealing with constraints is well established and the basic
feature of this approach is that gauge and reparametrization symmetries are
related with first-class constraints, not present in this model.  In
\cite{em}, the (hidden) SL(2,R) symmetry was found following
a somewhat modified Dirac's method, but their solution was criticized
in \cite{barcelos}
because they did not find the (supposedly also hidden) first-class
constraints that generate this symmetry.  This task has been pursued in
\cite{barcelos} more recently.  Working with the theory compactified in
the (light-cone) space-dimension ($x^-$) to a circle, the author was able
to identify among the
(infinite) set of second-class constraints, three apparently first-class
constraints.  He identified these constraints as the zero-modes of the
set of second-class
constraints of the gauge fixed theory.  Unfortunately these
constraints do not seem do correspond
to the generators of the SL(2,R) symmetry, as mentioned by the author.
This happens because these zero-mode
constraints are in fact second-class, which explains why they fail
to generate the required symmetry.

An alternative approach to the
problem has appeared in the literature quite recently \cite{ghosh}.  In
this work, the authors were able to clearly identify the generators of
this hidden symmetry as improper gauge constraints, as proposed by Regge
and Teitelboim \cite{rt,tcb,lledo}.  The separation of first-class
constraints into proper and improper was proposed in \cite{rt} in order to
resolve some difficulties of Dirac's method:  sometimes, among the
second-class constraints there appears to exist a subset of first-class
constraints.  In \cite{tcb} it was shown that a precise mathematical
treatment of this subset of first-class constraints reveals that they are
in fact second-class and, if treated as such, Dirac's procedure may be
consistently applied.  Proper gauge transformations then, represent true
symmetries of the theory and do not change the physical space of states.
Improper gauge transformations, as defined in \cite{rt,tcb} do change the
physical state of the system, mapping one physical solution into another
different one.  While proper gauge transformations can be eliminated by
gauge fixing, improper gauge symmetries cannot since that would exclude
physically allowed solutions.  Thus improper gauge transformations will
remain as hidden symmetries.

The outline of this paper is as follows.  In section II we review the FJ
formalism for invariant theories and show that the explicit symmetries, as
well as the hidden, are generated by the null eigenvectors of the
sympletic matrix, giving in this way an unified treatment for the subject
under this approach.  We end up this section with a discussion of two
simple examples of physical systems presenting hidden symmetries.  We show
next, in Section III, that the sympletic matrix for the Polyakov model in
the light cone gauge is singular and possess three null eigenvectors that
happen to be the generators of the SL(2,R) residual symmetry which,
therefore, becomes manifest in the Faddeev-Jackiw formalism.

\section{Faddeev-Jackiw Formalism For Invariant Theories}

Faddeev and Jackiw\cite{fadjack} showed that the treatment
of a Lagrangian that is
first-order in time derivatives dispense the use Dirac's formalism.
Consider a dynamical system with N bosonic degrees of freedom $q_i$
described by the following Lagrangian\footnote{We only consider a finite
number of bosonic coordinates, the extension to the fermionic case
\cite{govaertz} and to
field theory being straightforward\cite{bw1}.}

\begin{equation}
\label{fol}
L=a_i(q) \dot q_i - V(q) \; ; \;\;\;\; i=1,...,N
\end{equation}

\noindent $V(q)$ is the sympletic potential and $a_i(q)$ are the
components of the canonical one-form $a(q)=a_i(q) dq_i$.  This
presentation of the theory is called non-standard since its equations of
motion do not involve accelerations, and read

\begin{equation}
\label{eom}
\bar f_{ij} \dot q_j - {{\partial V} \over {\partial q_i}}=0
\end{equation}

\noindent with the sympletic matrix $\bar f_{ij}$ being defined by

\begin{equation}
\label{sm}
\bar f_{ij} = {{\partial a_j}\over{\partial q_i}} -
{{\partial a_i}\over{\partial q_j}}
\end{equation}

\noindent If this matrix is nonsingular, its inverse elements give
the Dirac brackets of the theory (\ref{fol}), with the dynamics
being described by

\begin{equation}
\label{db}
\{F,G\}={{\partial F}\over{\partial q_i}} (\bar f_{ij})^{-1}
{{\partial G}\over{\partial q_j}}
\end{equation}

\noindent However, when $\bar f_{ij}$ is singular the scheme
above no longer works due to the existence of true constraints that
reduce the number of independent degrees of freedom of the theory.  These
(sympletic) constraints will appear in this formalism as algebraic
relations $\Omega^a(q)$ needed to maintain the consistency of the
equations of motion upon multiplication by the (left) zero-modes
$v_i^a$ of $\bar f_{ij}$.  Then, from (\ref{eom}) we get

\begin{equation}
\label{tc}
\Omega^a \equiv (v_i^a)^T {{\partial V}\over{\partial q_i}}=0
\end{equation}

\noindent where $T$ stands for matrix transposition, and
\footnote{For the case of field theory, where the zero-modes are in
general operator valued objects, transposition is defined in the sense of
$\int (v^T A)B \equiv \int A(vB)$.}

\begin{equation}
\label{zm}
(v_i^a)^T \bar f_{ij} =0\; ; \;\;\;\; a=1,...,M
\end{equation}

\noindent define the M independent (left) zero-modes of $\bar f_{ij}$.

There are basically three situations that can arise out of the
{\it scalar product}
defining the sympletic constraints (\ref{tc}), which depends uniquely on
the {\it angle} between $v_i^{(a)}$ and
${{\partial V}\over{\partial q_i}}$ for each value of $a$.  In the
case of noninvariant systems
all zero-modes are nonorthogonal to the gradient of the sympletic
potential and constitute a true set of (sympletic) constraints that can be
iteratively implemented into the canonical sector of the original theory
(\ref{fol}), producing, step by step, a deformation of the original
sympletic matrix, which end up being nonsingular\cite{bw2}.
In Dirac's terminology,
this corresponds to a theory with purely second-class constraints.
The sympletic algorithm described above has been shown to lead to the
same results as that of Dirac's, although in a more economical way.
A completely different situation appears when the gradient of the
sympletic potential happens to be orthogonal to all zero-modes.  Then
condition (\ref{tc}) vanishes identically, the equations of motion are
automatically validated and no sympletic constraints appear.  This happens
due to symmetries in the sympletic potential, a situation typical of gauge
and reparametrization invariant theories.  There is no hope to deform the
sympletic
matrix into an invertible one unless a gauge-fixing term is introduced to
break the symmetry\cite{mw}, or one has the constraint solved and pass,
via Darboux's transformation, to a reduced canonical set\cite{tears}.
  Finally, we can have a mixed situation where
some zero-modes are orthogonal to ${{\partial V}\over{\partial q_i}}$ and
some are not.  In such  cases one uses the sympletic algorithm to eliminate
the constraints until only those zero-modes associated with symmetries
remain.

Let us assume, for definiteness, that all sympletic constraints have been
eliminated, and therefore only the zero-modes associated with gauge
symmetries are still present.  In first-order such theory can be written as

\begin{equation}
\label{gil}
L=a_k(q) \dot q_k + \dot\eta_a \Omega^a(q)- V(q)
\end{equation}

\noindent where the sympletic variables $q_k ; k=1,\ldots,N$
and $\eta_a ; a=1,\ldots , M$ form a set of gauge fields.
We have also relabeled the Lagrange multipliers $\lambda_a \rightarrow
-\dot\eta_a$\cite{mw}.  By hypothesis, the sympletic matrix
($\bar f_{km}$) constructed out of the $q_k$-variables only is
nonsingular, i.e., $\det \bar f \neq 0$.
In terms of the $q$'s and $\eta$'s the sympletic matrix
becomes,

\begin{equation}
\label{gism}
f=\left(
\begin{array}{cc}
(\bar f) & ({{\partial \Omega}\over{\partial q}})\\
{-({{\partial \Omega}\over{\partial q}})^T} & 0
\end{array}
\right)
\end{equation}

\noindent In this compact notation $(\bar f)$ represents a (NxN)
nonsingular matrix defined in (\ref{sm}).
$({{\partial \Omega}\over{\partial q}})$ represents a rectangular (NxM)
matrix defined as

\begin{equation}
\label{cm}
\left({{\partial \Omega}\over{\partial q}}\right)_{ka}= \left(
\begin{array}{cccc}
{{\partial \Omega^1}\over{\partial q_1}} &
{{\partial \Omega^2}\over{\partial q_1}} & {\ldots} &
{{\partial \Omega^M}\over{\partial q_1}} \\
{{\partial \Omega^1}\over{\partial q_2}} &
{{\partial \Omega^2}\over{\partial q_2}} & {\ldots} &
{{\partial \Omega^M}\over{\partial q_2}} \\
{\ldots} & {\ldots} & {\ldots} & {\ldots} \\
{{\partial \Omega^1}\over{\partial q_N}} &
{{\partial \Omega^2}\over{\partial q_N}} & {\ldots} &
{{\partial \Omega^M}\over{\partial q_N}} \\
\end{array}
\right)
\end{equation}

\noindent The sympletic matrix $(f)$ in (\ref{gism}) may have M zero-modes
with the following (block) structure

\begin{equation}
\label{gizm}
(v^a_k) = \left(
\begin{array}{cc}
-(\bar f_{km})^{-1}{{\partial \Omega^a}\over{\partial q_m}} \\
1^{(a)}
\end{array}
\right)
\end{equation}

\noindent where the first {\it element} is an (Nx1) column and $1^{(a)}$
is an (Mx1) column of zeros except in its a-th entry that is unity.  For
instance,

\begin{equation}
\label{column}
1^{(2)}= \left(
\begin{array}{c}
0 \\
1 \\
0 \\
\vdots \\
0
\end{array}
\right)
\end{equation}

\noindent Using (\ref{gism}) and (\ref{gizm}), the zero-mode condition
(\ref{zm}) can be written as two sets of equations; the first being
automatically satisfied, while the second set is valid, as long as
$\Omega^a$ defines an algebra under $\bar f$, i.e.,

\begin{equation}
\label{fcca}
\{\Omega^a,\Omega^b\}=C^{abc}\Omega^c
\end{equation}

\noindent with the curly bracket operation being defined by (\ref{db}).
Therefore,
the vectors $(v^a_k)$ in (\ref{gizm}) define a full set of
null eigenvectors of the sympletic matrix $(f)$ over
the constraint surface.  Since by hypothesis the gradient of the
sympletic potential is orthogonal to all zero-modes, they must be the
generators of the symmetry transformation that leave the action
invariant.  Specifically, we must have that the symmetry of the
action over the constraint surface reads\cite{mw}

\begin{eqnarray}
\label{se}
\delta_{\epsilon} q_m & = & -(\bar f_{mn})^{-1}
{{\partial\Omega^a}\over{\partial q_n}} \epsilon_a \nonumber \\
\delta_{\epsilon}\eta_a & = & -\epsilon_a
\end{eqnarray}

\noindent The $\epsilon_a$ form a set of infinitesimal parameters that
characterize the transformations.

The point to be stressed here is that the symmetry transformations
(\ref{se}) may reflect either the gauge or reparametrization properties of
an invariant theory, or they may as well be hidden symmetries of a
noninvariant model.  While the former is well described by Dirac's
formalism using the first-class constraints as symmetry generators, the
later has not an easy description in the usual canonical approach.

To give  simple but nontrivial illustrations of how the sympletic
formalism just described reveals a hidden symmetry, let us consider the
examples of Floreanini-Jackiw chiral boson\cite{florjack}, and 2D Maxwell
fields.  The theory of self-dual bosonic fields proposed by Floreanini and
Jackiw is described by the following action

\begin{equation}
\label{chiral}
S=\int d^2x (\dot\phi\phi' - \phi'\phi')
\end{equation}

\noindent where dot and prime have their usual meaning as time and space
derivatives.  The equations of motion possess a left mover field solution, as
it should, but in addition to that there is a space independent freedom in
terms of an arbitrary function of time $h(t)$ as

\begin{equation}
\label{symmetry}
\phi(x,t)=\phi_+(x+t) + h(t)
\end{equation}

\noindent In terms of Dirac's formalism, this theory has only a single
second-class constraint, being therefore unable to disclose the origin of
the symmetry (\ref{symmetry}).  The only hint for its presence comes from
the fact that Dirac's constraint matrix does not have a unique
inverse.  On the other hand, the model's sympletic matrix,

\begin{equation}
\label{matrix}
f(x-y)=2 \partial_x \delta (x-y)
\end{equation}

\noindent has as its only zero-mode, an arbitrary function of time which,
according to (\ref{se}), accounts for the symmetry (\ref{symmetry}) above.
To fix this invariance one has to adjust the boundary conditions such that
the sympletic matrix zero eigenvalue equation has no nontrivial solutions
which, in this case can be obtained with chiral boundary conditions.

Another simple but nontrivial example is given by Maxwell electrodynamics in
(1+1) space-time dimensions\cite{coleman}.  Following Dirac's approach one is
bound to find two first-class constraints

\begin{eqnarray}
\label{vinculos}
\Omega_0 & = & \pi_0 \nonumber\\
\Omega_1 & = & \partial_x E
\end{eqnarray}

\noindent which are respectively the primary constraint and the secondary
Gauss law constraint.  Introducing the gauge fixing conditions

\begin{eqnarray}
\label{gauge}
\chi_0 & = & A_0 \nonumber\\
\chi_1 & = & \partial_x A
\end{eqnarray}

\noindent corresponding to the radiation gauge, the constraints above will
form a second-class set.  If the boundary conditions of the problem are
chosen such that the Laplacian operator ($\partial_x^2$) has no nontrivial
zero-eigenvalue solution, then the Dirac matrix of constraints will have a
well defined and unique inverse.  Certainly in such a gauge all Dirac
brackets vanish identically, indicating that no propagating modes survive
after gauge fixing.  This result could have been told in advance since the
original phase-space having dimension four would result in a theory with
zero degrees of freedom after four gauge conditions have been imposed.
However, if the boundary conditions are less restrictive, there will be
some (hidden) residual symmetry left over which, as we will see, is
responsible for the theta-vacua structure displayed by the model.  Let us
consider the physical system above enclosed in a finite space-time
box\cite{pervushin}:

\begin{equation}
\label{box}
S=\int_{-{T\over 2}}^{{T\over 2}} dt \int_{-{R\over 2}}^{{R\over 2}}
dx \left( -{ 1 \over 4 }F_{\mu\nu}F^{\mu\nu}\right)
\end{equation}

\noindent The sympletic matrix is easily computed to be

\begin{equation}
\label{simpletica}
f(x-y)=\left(
\begin{array}{ccc}
0 & {-1} & 0 \\
1 & 0 & {-\partial_x}\\
0 & {-\partial_x} & 0
\end{array}
\right)
\delta(x-y)
\end{equation}

\noindent Here the sympletic variables are $\xi_k=\{A,E,\lambda\}$ with
$A=A_1$, $A_0=\dot\lambda$ and $E=F_{01}$. From (\ref{gism}) and
(\ref{gizm}) we get the zero-mode of (\ref{simpletica}) as

\begin{equation}
\label{modo-zero}
v(x)=\left(
\begin{array}{c}
\partial_x u\\
0\\
u
\end{array}
\right)
\end{equation}

\noindent with $u(x,t)$ being a totally arbitrary function of the space-time
coordinates.  This zero-mode displays the well known Abelian gauge symmetry
of the model which reads

\begin{eqnarray}
\delta A & = & \partial_x u \nonumber \\
\delta E & = & 0 \nonumber\\
\delta A_0 & = & \partial_t u
\end{eqnarray}

\noindent After imposing the radiation gauge, the sympletic matrix assumes the
form

\begin{equation}
f_{rad}(x-y)=\left(
\begin{array}{cccc}
0 & {-1} & 0 & \partial_x \\
1 & 0 & {-\partial_x} & 0 \\
0 & {-\partial_x} & 0  & 0 \\
\partial_x & 0 & 0 & 0
\end{array}
\right)
\delta(x-y)
\end{equation}

\noindent This matrix displays the structure advanced in (\ref{gism}) with

\begin{equation}
\bar f(x-y)=\left(
\begin{array}{cc}
0 & {-1} \\
1 & 0
\end{array}
\right)
\delta(x-y)
\end{equation}

\noindent and

\begin{equation}
\left({{\partial \Omega}\over{\partial q}}\right)=
\left(
\begin{array}{cc}
0 & \partial_x \\
-\partial_x & 0
\end{array}
\right)
\delta(x-y)
\end{equation}

\noindent and possess two zero-modes whose form is given by (\ref{gizm}).
For instance,

\begin{equation}
v^{(1)}(x)=\left(
\begin{array}{c}
\partial_x u\\
0\\
u\\
0
\end{array}
\right)
\end{equation}

\noindent with a similar one for the second zero-mode.  As discussed
above, the existence of a sympletic matrix zero-mode in this case will
depend on the Laplacian equation $\partial_x^2 u =0$ possessing a
nontrivial solution.  If the boundary conditions are chosen such that this
is the case, then we have that the two Laplacian eigenvectors

\begin{eqnarray}
u_1 & = & c_1(t) \nonumber\\
u_2 & = & c_2(t) x
\end{eqnarray}

\noindent do satisfy the following operator condition

\begin{equation}
D^{(k)} u_k = 0\; ;\;\; \mbox{(no sum)}
\end{equation}

\noindent with

\begin{equation}
D^{(k)}=\{\partial_x , x\partial_x -1\}
\end{equation}

\noindent It is easy to check that these operators close an algebra under
the commutator operation as

\begin{equation}
[D^{(k)},D^{(m)}]=C_n^{km} D^{(n)}
\end{equation}

\noindent where the only nonvanishing structure constants are
$C^{12}_1=C^{21}_1=1$.  One can see that redefining these operators as

\begin{eqnarray}
D^{(1)}\rightarrow \tilde D^{(1)} & = & D^{(1)}\nonumber\\
D^{(2)}\rightarrow \tilde D^{(2)} & = & {D^{(2)} \over D^{(1)}}
\end{eqnarray}

\noindent the set $\tilde D^{(k)}$ will form a simple representation
of the Heisenberg-Weyl algebra.

To see the physical consequences of this residual symmetry we follow
Ref.\cite{pervushin} closely.  Recall that the solution for the Laplacian
equation reads

\begin{equation}
u(x,t)=a(t)x+b(t)
\end{equation}

\noindent If one chooses $u(t,{R\over 2})=u(t,-{R\over 2})=0$, then $u(x,t)$
vanishes identically and no symmetry is obtained.  However, if one chooses
instead  $u(t,{R\over 2}) - u(t,-{R\over 2})= 2\pi n$, then $u(t,x)$ is
nontrivial and is given by

\begin{eqnarray}
u(t,x) & = & 2\pi{x\over R}N(t)\nonumber\\
N(t) & = & (n_+ - n_-){t\over T} +{1\over 2}(n_+ + n_-)
\end{eqnarray}

\noindent with $n_{\pm}$ being defined at the time boundary by

\begin{equation}
u_{\pm}(x)\equiv u(x,\pm{T\over 2})={{2\pi x}\over R} n_{\pm}
\end{equation}

\noindent In \cite{pervushin}, the authors have called this solution at
the time boundary as ``classical vacua''.  The zero-mode of the sympletic
matrix therefore describes the interpolation between two topologically
different vacua, $u_+(x)$ and $u_-(x)$, with relative winding number $\nu
= n_+ - n_-$.  Observe that the action for the Maxwell field now reduces
to that of a particle moving on a circle

\begin{equation}
S=\int_{-{T\over 2}}^{{T\over 2}} dt
{M\over 2} \dot N^2 = {1\over 2} M{(n_+ -n_-)^2\over T}
\end{equation}

\noindent with the particle's mass parameter being proportional to the
circle's inverse curvature $M={2\pi \over R}$.  Here

\begin{equation}
\nu=n_+ -n_- ={1\over{2\pi}} \int_{-{R\over 2}}^{{R\over 2}}dx
\int_{-{T\over 2}}^{{T\over 2}} dt F_{01}
\end{equation}

\noindent is the Pontryagin index and the zero-mode can safely be called
as the instanton solution.

\section{2D Induced Gravity}

In 2D there is no Einstein equation since the Hilbert-Einstein action
is a topological invariant measuring the genus of the manifold on
which one integrates.  Nevertheless, conformally invariant matter
fields induce an action over the metric fields through the conformal
anomaly. Let us now consider the action for the induced 2D gravity
\cite{polyakov}

\begin{equation}
\label{i2dg}
S =-k \int \sqrt{-g} R \Box^{-1} R d \tau d \sigma
\end{equation}

\noindent with $k=c/96\pi$, and $c$ being the central charge of the
matter field.  This effective action is nonlocal in the metric fields
but the nonlocality can be removed introducing
an auxiliary scalar field $\phi(\tau,\sigma)$\cite{polyakov,zadra,lina}

\begin{equation}
\label{lf}
S=- {1\over 2} \int \sqrt{-g} \left(g^{\alpha\beta} \partial_\alpha
\phi\; \partial_\beta\phi + \alpha R\phi\right)d\tau d\sigma
\end{equation}

\noindent where $\alpha$ is a constant parameter.  The coupling of
the scalar curvature $R$ with the auxiliary scalar field $\phi$
breaks the conformal invariance of the nonlocal model.  To break the
reparametrization invariance we must impose some gauge conditions.
We choose to work in the light cone gauge characterized by
$g_{\mu\nu}=\{g_{++}=2h ;\; g_{+-}=-1; \; g_{--}=0\}$.
With this choice the scalar curvature becomes

\begin{equation}
\label{sc}
R=-2\partial_{x^-}^2 h
\end{equation}

\noindent Using these results in (\ref{lf}) we obtain, after two
integrations by parts

\begin{equation}
\label{lcgl}
L=\int d x^- \left[\partial_{x^+}\phi \; \partial_{x^-}\phi +
h(\partial_{x^-}\phi \; \partial_{x^-}\phi +\alpha
\partial_{x^-}^2\phi)\right]
\end{equation}

As mentioned above, when described in
instant-front variables\cite{dirac2}, this model presents no constraints,
while in
light-cone variables only second-class constraints are present.  This
reflects the fact that all symmetries of the model have already been
fixed.  So, under Dirac's approach, this model should present no
symmetries.  To examine the existence of a (hidden) symmetry under the
Faddeev-Jackiw point of view we have to construct the sympletic matrix for
this model.  To this end we first relabel the Lagrange multiplier field as
$h\rightarrow\partial_{x^+}\lambda$, so that

\begin{equation}
\label{lcgl2}
L= \int d x^- \left[\partial_{x^+} \phi \; \partial_{x^-}
\phi + \partial_{x^+}
\lambda
( \partial_{x^-}\phi \; \partial_{x^-}\phi +\alpha \partial^2_{x^-}\phi)\right]
\end{equation}

\noindent The sympletic matrix (in terms of $\phi$ and $\lambda$) reads

\begin{equation}
\label{i2sm}
f(x,y)= \left(
\begin{array}{cc}
-2\partial_{x^-} & \alpha\partial_{x^-}^2 - 2\partial_{x^-}^2\phi - 2
\partial_{x^-}\phi\partial_{x^-} \\
-\alpha\partial_{x^-}^2 - 2\partial_{x^-}\phi\partial_{x^-} & 0
\end{array}
\right) \delta(x-y)
\end{equation}

\noindent One immediately notices the similar structure of this matrix
with that in (\ref{gism}).  Therefore, according to the discussion above,
if there exist solutions to the matrix equation

\begin{equation}
\label{zme}
\int d y^- f(x,y)v^{(a)}(y)  =0 \; ; \;\;\;\; a=1,\ldots,M
\end{equation}

\noindent then they must corresponds to the zero-modes of $f(x,y)$ that
will reveal the symmetry remaining in
the model.  According to (\ref{gizm}) the zero-modes must have the
following structure

\begin{equation}
\label{zme2}
v^{(a)}(y)= \left(
\begin{array}{c}
{1\over 4}\int dz^-\epsilon(y^- -z^-)\left[\alpha\partial_{z^-}^2g_a(z)- 2
\partial_{z^-}(g_a(z)\partial_{z^-}\phi )\right] \\
g_a(y)
\end{array}
\right)
\end{equation}

\noindent which, after simple algebra, can be written in the following
form

\begin{equation}
\label{zme3}
v^{(a)}(y)= \left(
\begin{array}{c}
{1\over 2}\alpha\partial_{y^-}g_a(y)- g_a(y)\partial_{y^-}\phi \\
g_a(y)
\end{array}
\right)
\end{equation}

\noindent with $g_a(x)$ being a set of some yet undetermined functions.  In
components, the matrix equation (\ref{zme}) becomes the following pair of
equations

\begin{eqnarray}
\label{zme4}
0&=&\int dy^-\partial_{x^-}\delta(x^- - y^-)\left(\alpha \partial_{y^-}
g_a(y) - 2 g_a(y) \partial_{y^-}\phi\right) \nonumber\\
& &\mbox{} + \int dx^-\left(\alpha \partial_{x^-}^2 - 2
\partial_{x^-}^2\phi - 2
\partial_{x^-}\phi\partial_{x^-}\right)\delta(x^- - y^-)g_a(y) \nonumber\\
0&=& \int dy^-\left(-2 \partial_{x^-}\phi\partial_{x^-}+\alpha
\partial_{y^-}^2\right)
\delta(x^- - y^-)
\left({\alpha\over 2} \partial_{y^-}g_a(y) - g_a(y)
\partial_{y^-}\phi\right)\nonumber
\end{eqnarray}

\noindent It is a simple algebra to check that the first equation is
trivially satisfied, therefore imposing no restriction over the functional
form of $g_a(x)$.  The second equation, on the other hand, is only
satisfied if

\begin{equation}
\alpha^2 \partial_{x^-}^3 g_a(x)=0~; \alpha\neq 0
\end{equation}

\noindent There are therefore three distinct null eigenvalues for the
sympletic matrix (\ref{i2sm}) that reads

\begin{eqnarray}
g_1(x) & = & \epsilon^{(1)}(x^+)\nonumber\\
g_2(x) & = & \epsilon^{(2)}(x^+)x^-\nonumber\\
g_3(x) & = & \epsilon^{(3)}(x^+)(x^-)^2
\end{eqnarray}

\noindent If one removes these three zero-modes (with properly chosen
boundary conditions) then the sympletic matrix becomes nonsingular and the
elements of the inverse  give the Dirac brackets that quantize the theory.
Observe that the three differential operators satisfying

\begin{equation}
D^{(a)} g_a=0~;~\mbox{(no sum)}
\end{equation}

\noindent are given by

\begin{equation}
D^{(a)}=\{ \partial_{x^-};\; x^- \partial_{x^-} - 1;\; (x^-)^2
\partial_{x^-}- 2 x^- \}
\end{equation}

\noindent that can be easily verified to be  the generators of the
SL(2,R) group algebra.  Also, from
(\ref{se}) we obtain the symmetry transformation of the scalar field under
this group of transformations as

\begin{eqnarray}
\delta_{\epsilon}\phi & = & -\left(\epsilon^{(1)}(x^+)+ x^-
\epsilon^{(2)}(x^+)+(x^-)^2
\epsilon^{(3)}(x^+)\right)\partial_{x^-}\phi \nonumber\\
& &\mbox{} + {\alpha\over 2} \left( \epsilon^{(2)}(x^+) + 2 x^-
\epsilon^{(3)}(x^+)\right)
\end{eqnarray}

In conclusion, under the geometric approach of Faddeev-Jackiw, the SL(2,R)
residual symmetry of Polyakov's model manifest itself in terms of the
zero-modes of the sympletic matrix, just as any other explicit
symmetry, such as gauge and reparametrization invariances.  While the
study of symmetries inside Dirac's formalism is based on the existence of
first-class constraints, in the Faddeev-Jackiw formalism it relies on the
orthogonality of the sympletic matrix zero-modes with the gradient of
the (sympletic) potential.  These (null) eigenvectors generate thus the
isopotential lines of the theory, i.e. its symmetry lines, bypassing the
first-class constraints for doing this job.
This characterization of the Faddeev-Jackiw
method in terms of null eigenvectors of the sympletic matrix, instead
of constraints seems to be a definite advantage of this
method over Dirac's formalism since it avoids the use of first-class
constraints, which in cases as the one above are certainly not
avaliable.\vspace{0.1cm}\\

\noindent ACKNOWLEDGMENTS:  We want to thank Drs. J.C.Brunelli and A.Das for
many useful suggestions and comments.\vspace{0.1cm}\\

\noindent Note Added:  After the completion of our work we became aware of
the Ref.\cite{shat,rai,delius} where these authors were able to recover
Polyakov's 2D gravity using the coadjoint orbit method.  In particular
Alekseev and Shatashvili\cite{shat} have shown the existence of a SL(2,R)
symmetry in the 2D gravity by relaxing the periodicity boundary condition.
The main point of this method is to exploit the fact that a coadjoint
orbit of a group admits a sympletic two-form that may be used to construct
an action that enjoys at least as much symmetry as the group whose
coadjoint orbit is under consideration.  It should be noticed that the
method of our paper, using the Faddeev-Jackiw formalism, follows the
opposite pathway, in the sense that we use the zero-modes of a singular
sympletic two-form to determine the symmetries enjoyed by the action that
produced that particular singular sympletic form.  It is also worth to
point out that the coadjoint orbit method is a highy mathematical subject
while our method is acessible to all.

\end{document}